\documentclass{article}
\usepackage[utf8]{inputenc}
\usepackage{units}
\usepackage{blindtext}
\usepackage{graphicx}
\usepackage{kpfonts}
\usepackage{amsmath}
\usepackage[a4paper, total={6in, 9in}]{geometry}
\usepackage{authblk}
\usepackage{comment}
\usepackage{color}
\usepackage{our}
\usepackage{setspace}
\usepackage{cleveref}

\begin{document}

\title{Strong-Field QED Experiments using the BELLA PW Laser Dual Beamlines}

\author[1,e1]{M. Turner \thanks{corresponding author}}
\author[1,e2]{S.~S. Bulanov }
\author[1,e3]{C. Benedetti }
\author[1,e4]{A.~J. Gonsalves }
\author[2,e5]{W.~P. Leemans }
\author[1,e6]{K. Nakamura }
\author[1,e7]{J. van Tilborg }
\author[1,e8]{C.~B. Schroeder}
\author[1,e9]{C.~G.~R. Geddes}
\author[1,e10]{E. Esarey }

\affil[1]{Lawrence Berkeley National Laboratory, Berkeley, USA}
\affil[2]{Deutsches Elektronen-Synchrotron DESY, Hamburg, Germany}

\affil[e1]{marleneturner@lbl.gov}
\affil[e2]{sbulanov@lbl.gov}
\affil[e3]{cbenedetti@lbl.gov}
\affil[e4]{ajgonsalves@lbl.gov}
\affil[e5]{wim.leemans@desy.de}
\affil[e6]{knakamura@lbl.gov}
\affil[e7]{jvantilborg@lbl.gov}
\affil[e8]{cbSchroeder@lbl.gov}
\affil[e9]{cgrgeddes@lbl.gov}
\affil[e10]{ehesarey@lbl.gov}

\date{March 2022}

\maketitle

\begin{abstract}

The Petawatt (PW) laser facility of the Berkeley Lab Laser Accelerator (BELLA) Center has recently commissioned its second laser pulse transport line. This new beamline can be operated in parallel with the first beamline and enables strong-field quantum electrodynamics (SF-QED) experiments at BELLA. In this paper, we present an overview of the upgraded BELLA PW facility with a SF-QED experimental layout in which intense laser pulses collide with GeV-class laser-wakefield-accelerated electron beams. We present simulation results showing that experiments will allow the study of laser-particle interactions from the classical to the SF-QED regime with a nonlinear quantum parameter of up to $\chi\sim$2. In addition, we show that experiments will enable the study and production of GeV-class, mrad-divergence positron beams via the Breit-Wheeler process.
\end{abstract}

\section{Introduction}
\label{sec:Intro}

Classical and quantum electrodynamics have been extensively and successfully verified for almost all parameter ranges. However, open questions remain for interactions in \textbf{strong} electromagnetic (EM) fields \cite{piazza.rmp.2012,zhang.pop.2020,gonoskov.rmp.2022,fedotov.arxiv.2022}. For example, classical electrodynamics overestimates the radiation reaction (which is  what affects the dynamics of a radiating particle ) and allows for the emission of photons with energy greater than the particle energy, a problem that can be addressed by switching to the quantum description. Both open questions and potential applications motivate the study of strong field (SF) interactions in experiments to, e.g.:
\begin{enumerate}
    \item \textit{Develop an experimental framework} that provides a consistent way to verify theoretical and simulation predictions from the classical to the quantum electrodynamics (QED) regime, including linear and nonlinear effects as well as multi-staged processes typical of SF-QED environments. Previous experiments either operated in a parameter space where the nonlinear quantum parameter $\chi$ was clearly below $1$ \cite{bula.prl.1996,burke.prl.1997}, or provided a limited set of data \cite{cole.prx.2018,poder.prx.2018}. Due to the increased availability of high power lasers \cite{danson.hplse.2019}, multiple facilities (as detailed later in this section) are planning experiments to reach $\chi>1$ by using higher laser intensities, more energetic particle beams, and higher repetition rate lasers.
    
    \item \textit{Evaluate whether strong-field interactions may provide competitive $\gamma$-ray or positron sources}~\cite{gonoskov.rmp.2022}. Strong EM fields may be used to produce high-flux $\gamma$-rays (see, e.g., Refs.~\cite{gonoskov.prx.2017,magnusson.pra.2019,PhysRevLett.113.224801}) and low divergence positron sources~\cite{commphys,pos2}. Positron sources are possible bottlenecks for future TeV-class lepton colliders \cite{benedetti.arxiv.2022}. Understanding whether strong EM fields and QED effects can generate sources that compete with those used in conventional accelerators \cite{musumeci.arxiv.2022} is a high priority for the high energy physics community.
\end{enumerate}

The basic building blocks of SF-QED are the Compton effect (photon emission by an electron) and the Breit-Wheeler effect (photon decay into an electron-positron pair) in strong EM fields~\cite{ritus.jslr.1985}. It is most convenient to characterize these interactions in terms of Lorentz invariant parameters: 
\begin{align} 
\mathcal{F} & =(\vec{E}^2 - c^2 \vec{B}^2)/\Ecrit^2,\\
\mathcal{G} & =c \vec{B}\cdot\vec{E}/\Ecrit^2,\\ 
\chi_e & =\gamma \sqrt{(\vec{E} + \vec{v} \times \vec{B})^2 - (\vec{E}\cdot\vec{v}/c)^2}/\Ecrit, \\ 
\chi_\gamma & =(\hbar \omega/m c^2)\sqrt{(\vec{E} + \left(c^2 \vec{k}/\omega\right) \times \vec{B})^2 - \left(\vec{E}\cdot\left(c\vec{k}/\omega\right)\right)^2}/\Ecrit, 
\end{align} 
where $c$ is the speed of light, $\hbar$ is the Planck constant, and $m$ is the electron mass. Here, $\vec{E}$ and $\vec{B}$ are the electric and magnetic fields, respectively, and $\mathcal{F}$ and $\mathcal{G}$ are the Poincar\'e invariants of the EM field \cite{schwinger.pr.1951}.
%Denoting by $\vec{E}$ and $\vec{B}$ the electric and magnetic fields, $\mathcal{F}$ and $\mathcal{G}$ are the Poincar\'e invariants of the EM field \cite{schwinger.pr.1951}. 
The particle momentum is defined as $p^\mu = \gamma m (c, \vec{v})$, where $\gamma = (1 - \vec{v}^2/c^2)^{-1/2}$ and $\vec{v}$ is the particle velocity. The photon momentum is defined as $\hbar k^\mu = (\hbar \omega/c) (1, \vec{n})$, where $\omega$ is the photon frequency, $\vec{n}$ its propagation direction and the photon is on-shell ($k_\mu k_\mu=0$). Whereas $\mathcal{F}$ and $\mathcal{G}$ characterize the fields itself, $\chi_e$ and $\chi_\gamma$ characterize the interaction of charged particles (e.g., electrons) and photons, respectively, with the strong fields (we recall that an EM field is considered strong when it is of the order of the QED critical field \cite{schwinger.pr.1951,sauter.zp.1931,heisenberg.zp.1936}, \unit[$E_{crit}=1.32\times10^{18}$]{V/m} or \unit[$B_{crit}=4.41\times10^{9}$]{T}). All above defined Lorentz invariant parameters are normalized to $\Ecrit$, which provides a natural scale for the onset of quantum effects in the electromagnetic interactions (i.e., when $\mathcal{F}$, $\mathcal{G}$, $\chi_e$, $\chi_\gamma\sim1$).

Strong EM fields can be found in different environments, including in close proximity of compact astrophysical objects (such as magnetars and black holes) \cite{olausen.apjs.2014,crinquand.prl.2020}, high-Z nuclei \cite{reinhardt.rpp.1977}, dense particle beams (at the interaction point of high energy particle accelerators) \cite{yakimenko.prl.2019}, aligned crystals \cite{uggerhoj.rmp.2005}, and in the foci of high power lasers \cite{danson.hplse.2019}. Some of these environments provide fields of the order of the critical strength, but are not accessible in any laboratory in the foreseeable future. Others can reach the critical strength in the reference frame of a sufficiently high energy particle or in fixed plasma targets. At the current state-of-the-art, laboratory SF-QED experiments will require an interaction between energetic particles and EM fields ($\chi_e=\gamma E/\Ecrit$). 

%The typical value of the parameter $\chi_e$ for an electron interaction with an electric field is $\chi_e=\gamma E/\Ecrit$, which means that $\chi_e=1$ can be interpreted as that the EM field strength in the electron rest frame is equal to the critical one. For example, the earth magnetic field appears to be of critical strength as seen by a cosmic-ray electron with an energy exceeding \unit[$10^{19}$]{eV} \cite{intro3} or the electric field of a bunch at a future TeV electron-positron linear collider approaches the critical field in the frame of the oncoming bunch \cite{intro2}.%That is for example the case for aligned crystals and in the foci of high power lasers.

%A distinct class of interactions is possible with high-intensity lasers, which can, at sufficiently \cite{gonoskov.rmp.2022} large field strength, accelerate initially stationary electrons to sufficiently high energy to make radiation losses and quantum effects important. In this case the laser plays the role of both accelerator and target.

Previous experiments reached a maximum nonlinear quantum parameter (in the following denoted as $\chi_{e,max}$) of $\chi_{e,max}\sim0.3$ in the E144 experiment at SLAC~\cite{bula.prl.1996,burke.prl.1997}, and $\chi_{e,max}\sim0.2$ in the GEMINI experiment at CLF~\cite{cole.prx.2018,poder.prx.2018}. Experiments using aligned crystals were reported recently \cite{uggerhoj.rmp.2005,wistisen.ncomm.2018,wistisen.prr.2019}, % reaching %$\chi_{e,max}\sim\leq1.4$
but require specific analysis techniques and positron beams. Experiments using particle colliders are proposed \cite{yakimenko.prl.2019}, but are inaccessible due to the lack of accelerators with necessary parameters. Therefore, interactions of electrons with high intensity laser pulses provide the most promising immediate path to increase $\chi_{e}$ or $\chi_{\gamma}$ above unity.  %The main difference between the SLAC and CLF experiments was that the former used of the conventional SLAC accelerator to produce high energy electrons, while the latter resorted to an all-optical scheme where electron beams were produced via laser wakefield acceleration (LWFA).

On that path, SLAC is planning the E320 experiment, and DESY is planning the LUXE experiment \cite{abramowicz.arxiv.2021} using conventionally accelerated \unit[$10$ or $17.5$]{GeV} electron beams in collision with tens of TW laser pulses. The University of Michigan ZEUS facility will use two laser pulses (with \unit[2.5]{PW} and \unit[0.5]{PW}), one to accelerate electrons in a laser wakefield accelerator (LWFA) (to either \unit[$\gtrsim 10$]{GeV}, or several GeV) and one to provide the EM field (with intensity \unit[$10^{21}$]{W/cm$^{2}$}, or \unit[$10^{23}$]{W/cm$^{2}$}). Other laser facilities with active SF-QED study programs include J-Karen in Japan, Apollon in France, CORELS in Korea, CALA in Germany, ELI NP in Romania with interaction chambers with colliding \unit[10]{PW} laser pulses \cite{ELINP.rom.rep.2016,ELI.HPLSE.2016}, and ELI BL in Czech Republic, SEL in China \cite{wang202213} (for an expanded list see Ref. \cite{gonoskov.rmp.2022} and \cite{Pwlaser.HPLS.2019} for PW laser facilities).

In this paper, we assess the potential for SF-QED experiments at the BELLA Center of the Lawrence Berkeley National Laboratory. The BELLA Center hosts a \unit[1]{Hz}, petawatt (PW) laser facility called the BELLA PW, and has recently commissioned a second high-power laser  beamline (2BL) that enables SF-QED experiments. Simulation studies (see Sec.~\ref{sec:SFQEDsimres}) show that experiments on BELLA PW will allow to investigate a wide range of $\chi_{e}$ reaching immediately up to 2, and potentially up to 4 after optimizations, which is very attractive at the unique \unit[1]{Hz} repetition rate of the laser. Additionally, the BELLA Center experimental teams have many years of experience on laser operation and laser-driven plasma wakefield acceleration of electron beams \cite{leemans.nphys.2006,leemans.prl.2014,steinke.nature.2016, nakamura.ieee.2017,gonsalves.prl.2019}. 

This paper is organized as follows. Section~\ref{sec:facility} provides an overview of and general introduction to the BELLA PW facility, Sec.~\ref{sec:interactiongeom} discusses the two basic SF-QED laser-particle interaction geometries, Sec.~\ref{sec:ebeamsim} provides an overview of experimentally achievable electron beam parameters using the BELLA PW laser, Sec.~\ref{sec:SFQEDsimres} discusses the scientific reach of SF-QED experiments based on simulation results, Sec.~\ref{sec:expgeom} explores experimental layouts at BELLA PW, and Sec.~\ref{sec:Conclusions} closes with a summary and the conclusions.

\section{BELLA PW Experimental Facility and Dual Beamlines}
\label{sec:facility}

\begin{figure}[htb!]
    \centering
    \includegraphics[width=\columnwidth]{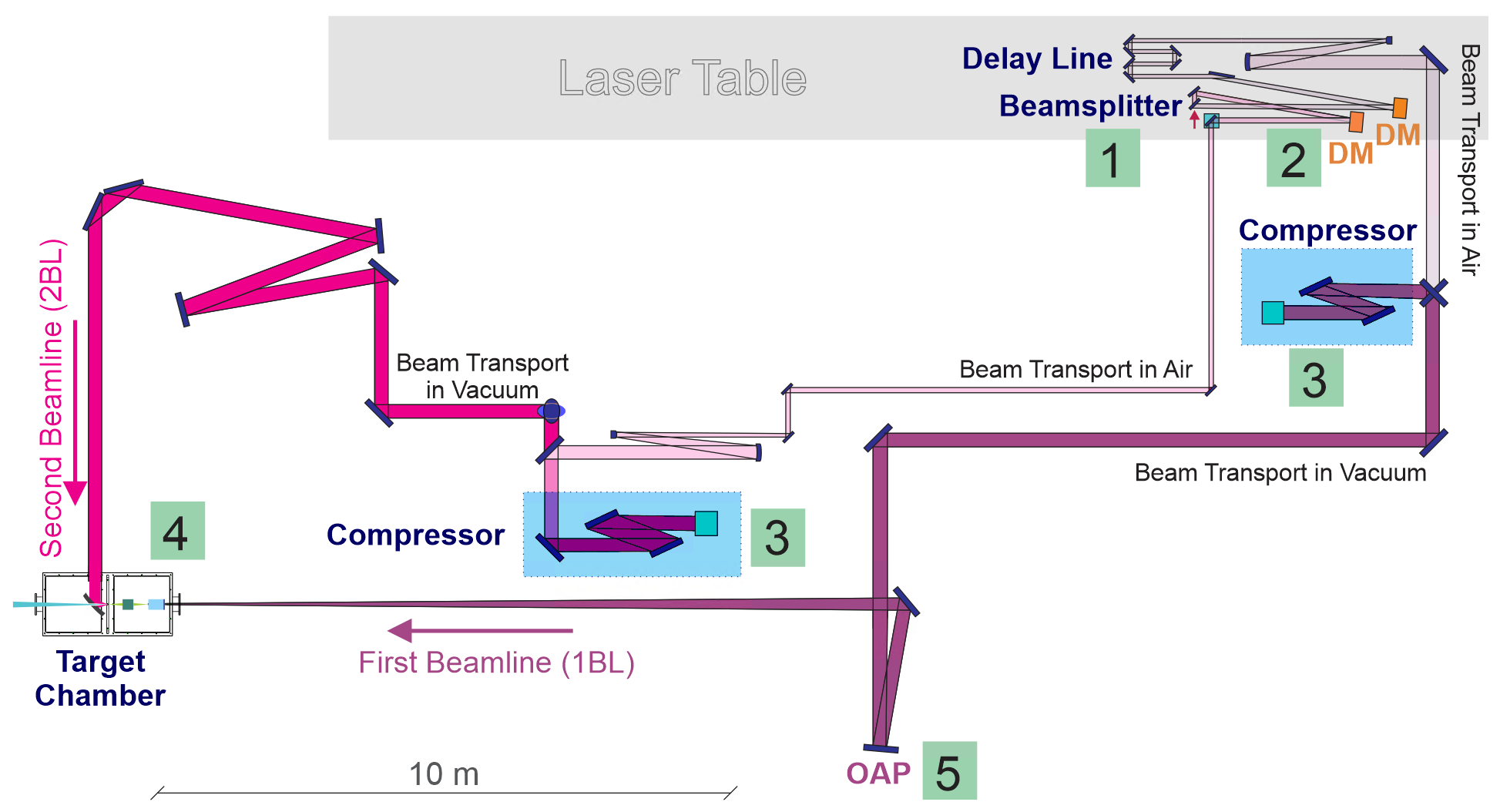}
    \caption{Schematic layout of the BELLA Petawatt dual beamlines for SF-QED experiments.}
    \label{fig:Layout}
\end{figure}

This section provides an overview of the BELLA PW facility, experimental parameters, and planned experiments. The facility comprises a petawatt laser system, three laser pulse transport lines (1BL, 2BL, and iP2 beamline) and two experimental target chambers. While all parts will be mentioned briefly, the focus will be on the components required for SF-QED experiments: the BELLA PW laser system, first and second beamline as well as their target chamber, which are illustrated in Fig.~\ref{fig:Layout}. 

The core of the BELLA PW facility is a petawatt-class laser system, which provides uncompressed pulses with a total energy up to \unit[60]{J} per pulse at \unit[1]{Hz} repetition rate. Pulses are transported from the laser table to the first target chamber via two pulse transport lines (see Fig.~\ref{fig:Layout}) named first (1BL) and second (2BL) beamline. Due to losses in the compressor and beamlines, a total of \unit[$\sim$40]{J} of pulse energy (or \unit[1.2]{PW} of maximum power) is available for experiments in the target chamber (see location {4} on Fig.~\ref{fig:Layout}). 

To send laser light into both beamlines, pulses are split after amplification and before compression by a beamsplitter on the laser table (see location [1] in Fig.~\ref{fig:Layout}). Pulse energy reflected by the beamsplitter is transported in the first beamline, the remaining energy is transmitted through the beamsplitter and transported in the second beamline. The choice of beamsplitter reflectivity defines the energy splitting ratio and the ratio can be adjusted by exchanging the optic. 

Both first and second beamline use a deformable mirror (see location [2] on Fig.~\ref{fig:Layout}) for beam shaping and beam profile optimization and a chirped-pulse-amplification compressor (see locations [3] on Fig.~\ref{fig:Layout}) for compression down to lengths of \unit[$\tau\sim$30-40]{fs}. The two beamlines share a Dazzler for spectral pulse shaping before compression. A delay line on the first beamline as well as a motorized stage inside the second beamline compressor allow for the adjustment of timing between first and second beamline pulses. 

First beamline was commissioned in 2012 together with the BELLA PW laser and has been operating successfully since. The pulse propagating in the first beamline is focused by a \unit[13.5]{m} focal length off-axis parabolic mirror (OAP) (see location [5] on Fig.~\ref{fig:Layout}) to a focal spot size of \unit[$w_{\mathrm{0}}=$53]{$\mu$m} inside the target chamber. Second beamline commissioning was completed in 2022 and allows for several focusing options, including \unit[f=6.5, 10.4, 13.5 and 18.0]{m} OAPs or a flat mirror to transport the collimated beam (up to a diameter of \unit[$\sim15$]{cm}) into the target chamber (see location [4] on Fig.~\ref{fig:Layout}). In Fig.~\ref{fig:Layout}, the layout using the flat mirror (suitable for SF-QED experiments) is shown.

Pulses with a diameter of up to \unit[15]{cm} can enter the current target chamber, limited by the location of the chamber support. A new target chamber is required to allow use of the full aperture beam (diameter \unit[$\sim20$]{cm}) and could also be designed to facilitate the challenging particle detection and required radiation shielding for SF-QED experiments.

Not shown in Fig.~\ref{fig:Layout} is the BELLA PW high intensity laser beamline (iP2) and the iP2 target chamber that is located downstream the target chamber in Fig.~\ref{fig:Layout}, as an extension to the first beamline. The iP2 extension uses the first beamline laser pulse that propagates through the target chamber shown in Fig.~\ref{fig:Layout} and provides a laser focus with an intensity of \unit[$>10^{21}$]{W/cm$^2$}, using a short-focal length OAP (f/2.5) and is used, e.g., for solid target experiments. However, the iP2 target chamber provides access for only one laser pulse and therefore it would require a significant effort to devise an interaction configuration, which is suitable for SF-QED experiments \cite{hakimi.pop.2022}.

\subsection{Dual Beamline Experiments Planned at BELLA PW}
\label{sec:plannedexp}
The new dual pulse capabilities of the upgraded BELLA PW facility enable a variety of unique experiments. Construction of the second beamline was motivated by plasma staging experiments~\cite{steinke.pop.2016} enabling research towards a high-energy physics particle collider at the energy frontier. The goal of these experiments is to demonstrate at the GeV-level that an electron beam accelerated in a LWFA stage can be further accelerated in a subsequent LWFA stage with high charge capture and high beam quality. For that, the first (second) beamline will be used to drive a wakefield in the first (second) plasma stage.

Additionally, the BELLA PW facility is planning to use the dual pulse capabilities for single stage development, e.g., to optimize energy gain. One idea is to guide drive pulses in conditioned hydrodynamic optically field-ionized plasma channels~\cite{shalloo.prab.2019, miao.prl.2020}. These experiments will use the 2BL pulse to optically field ionize gas and to heat the plasma, leading to the formation of a plasma channel via hydrodynamic expansion. The 1BL pulse will then drive high amplitude wakefields in that channel, which can be used to accelerate electrons to the \unit[$\gtrsim10$]{GeV}-level. 

Also enabled by the new second beamline, and a natural follow up to the two experiments mentioned above, are SF-QED studies, which are the topic of this article. The physics reach of potential experiments is discussed in Sec.~\ref{sec:SFQEDsimres} and an experimental layout is proposed in Sec.~\ref{sec:expgeom}.

\section{SF-QED Particle and Field Interaction Geometries}
\label{sec:interactiongeom}

This section discusses the motion of charged particles (here we consider electrons) in laser fields and in the presence of SF-QED effects. There are two typical interaction geometries: 1) relativistic electrons - laser and 2) laser - laser. Though it is well known that geometry 2) requires much higher laser intensities to observe radiation dominance and quantum effects than geometry 1), it is instructive to revisit both options to evaluate the thresholds for radiation reaction and the onset of quantum effects (in what follows we used formulae from Ref.~\cite{gonoskov.rmp.2022}), to emphasize the differences between the two interactions and to explain the experimental layout choice for experiments at BELLA PW, which is geometry 1).  

\textbf{1) Electron-laser interaction, ($||$).}  An electron with a relativistic factor $\gamma$ collides with a high intensity laser pulse with normalized strength $a_0$ ($a_0\simeq0.855\times 10^{-9}(I$[W/cm$^2$])$^{1/2}\lambda$[um], where $I$ is the laser pulse peak intensity and $\lambda$ the laser wavelength). When $\gamma\gg a_0$, the electron motion induced by the EM field is mostly \textit{longitudinal} (with respect to the laser propagation direction) and is relatively unaffected by the Lorentz force. The laser pulse forces the electron to radiate, as its EM field serves as a "target", resulting in electron energy loss and deceleration.

\textbf{2) Laser-laser interaction, ($\perp$).} An initially non-relativistic electron (e.g., from a plasma) interacts with the field of two colliding circularly polarized laser pulses (with combined intensity $a_0$). Here we consider a setup where the electron circulates in the anti-node of a standing light wave, which means that it experiences only the electric field. The electron motion is mostly \textit{transverse}, dominated by the Lorentz and radiation reaction forces. While the electron loses energy due to radiation emission, it is also being continuously re-accelerated by the Lorentz force in the electric field which compensates the energy loss. Here, the laser serves not only as a "target" but also as an accelerator.

The definition of the parameter $\chi_e$ for these two configurations ($\chi_e^{||}$, $\chi_e^\perp$) clarifies their similarities and differences:
\begin{align}
    \chi_e^{||}=\gamma\frac{E}{\Ecrit}\left(1-\beta\cos\theta_{pk}\right),
    \label{eq:chiparallel}
    \\
     \chi_e^\perp=\gamma\frac{E}{\Ecrit}\sqrt{1-\beta^2\cos^2\theta_{pe}},
     \label{eq:chiperp}
\end{align}
%Superscripts $||$ and $\perp$ stand for longitudinal and transverse geometries detailed above, 
where $\beta=v/c$ is the normalized electron velocity, $\theta_{pk}$ is the angle between the electron momentum and the wave vector, and $\theta_{pe}$ is the angle between the electron momentum and the field vector.

Both $\chi_e^{||}$ and $\chi_e^\perp$ are proportional to $\gamma E/\Ecrit$ or the EM field strength in the electron rest frame normalized to the critical field. However, their angular dependence is different. While $\chi_e^{||}$ is maximum for a head-on collision,  $\chi_e^\perp$ is maximum when the field direction and particle momentum are perpendicular, which corresponds to the particle motion under the action of the Lorentz force in a circularly polarized electric field. For ($||$) a head-on collision is assumed, i.e., $\theta_{pk}=\pi$.

%Though the goal of this paper is to show how SF-QED effects can be observed at the BELLA PW facility,
When interacting with a counter-propagating laser pulse, electron dynamics start to be affected by radiation emission before quantum effects come into play. This is usually accounted for by including the \textbf{radiation reaction} force into the classical equations of motion of an electron in an EM field \cite{review.mod.phys.2020}.  In what follows we will use the Landau-Lifshitz equation of motion \cite{landau.1981} to determine at which field strength and particle energy the radiation reaction effects start to dominate electron behavior for each of the two geometries.

For ($||$), the onset of the radiation dominated particle motion is usually defined as the loss of half of the initial electron energy by the end of the interaction. If we assume head-on collision, this is the case when the laser field strength is greater than:
\begin{equation}
    a_0^{||}>(\varepsilon_{rad}\omega_l \tau \gamma)^{-1/2}\simeq 82(\lambda[\mu \hbox{m}]/\mathcal{E}_e[\hbox{GeV}]N)^{1/2}.
\end{equation}
Here $\varepsilon_{rad}=(2\alpha/3)(\hbar\omega_l/mc^2)$, $\alpha$ is the fine structure constant, $\tau$ and $\omega_l$ are the laser duration and frequency, respectively, $N=c\tau/\lambda$ is the number of laser cycles, and $\mathcal{E}_e$ is initial electron energy. For ($\perp$), a different definition for the field strength that leads to the onset of radiation dominated particle motion is usually used, since electrons not only radiate their energy away but are also re-accelerated by the electromagnetic field. Thus, when an electron emits the same amount of energy as it gains from the field per cycle, the interaction enters the radiation dominated regime. This happens when the field strength is greater than: 
\begin{equation}
    a_0^\perp>\varepsilon_{rad}^{-1/3}\simeq 4.4\times 10^2\lambda^{-1/3}[\mu\hbox{m}].
\end{equation}
Here $\theta_{pe}=\pi/2$.

For example, for a \unit[5]{GeV} electron beam colliding with a \unit[800]{nm}, \unit[10]{cycle} laser pulse, the characteristic value of the laser field strengths for the ($||$) configuration is $a_0^{||}\simeq 10$, which corresponds to a peak intensity of $4\times 10^{20}$ W/cm$^2$. In the ($\perp$) case the characteristic value of the field strength is $a_0^\perp=474$, which corresponds to $3.5\times 10^{23}$ W/cm$^2$, almost three orders of magnitude higher.

The onset of \textbf{quantum effects} can be characterized by the values of electron energy and field strength, which results in an emission of a photon that can carry away almost all electron energy. It is estimated using the critical frequency of the classical radiation spectrum multiplied by $\hbar$ as a characteristic photon energy which is compared to the electron energy, or from the condition $\chi_e\sim 1$. This occurs when field strengths exceed:
\begin{align}
    a_0^{||} & >\frac{2\alpha}{3\varepsilon_{rad}\gamma}\simeq 205 \frac{\lambda[\mu \hbox{m}]}{\mathcal{E}_e[\hbox{GeV}]},
    \\
    a_0^\perp & >\frac{4\alpha^2}{9\varepsilon_{rad}}\simeq 2\times 10^3\lambda [\mu \hbox{m}]. \label{aperpQ}
\end{align}
For example, for a \unit[5]{GeV} electron beam colliding with a \unit[800]{nm} laser pulse, the characteristic value of the laser field strength for the ($||$) configuration is $a_0^{||}\simeq 34$, which corresponds to a peak intensity of $3.2\times 10^{21}$ W/cm$^2$, whereas for an electron moving in the focus of two colliding \unit[800]{nm} laser pulses the characteristic field strength is $a_0^{\perp}\simeq 1600$,
which corresponds to an intensity of $5.5\times 10^{24}$ W/cm$^2$, three orders of magnitude higher. The main reason for such a difference is that the  ($\perp$) configuration needs to both accelerate electrons to multi-GeV energies and provide the field component perpendicular to the electron momentum strong enough to lead to the high energy photon emission. This becomes increasingly difficult since in strong fields and in the presence of the radiation reaction, the electron momentum tends to align with the field vector direction \cite{bulanov.pre.2011} (see also \cite{esirkepov.pla.2017,gonoskov.pop.2018,ekman.njp.2021}), which is taken into account when deriving Eq. (\ref{aperpQ}). 

The BELLA PW laser system can provide pulse intensities up to \unit[1.4$\times10^{22}$]{W/cm$^2$} ($a_0\simeq 80$) in 2BL, or electron energies up to   \unit[$12.4$]{GeV} ($\gamma=24300$) in 1BL (see Secs.~\ref{sec:facility}, \ref{sec:ebeamsim}, and \ref{sec:expgeom} for details). Estimates in this section clarify that only the interaction of an electron beam with a counter-propagating laser pulse allows the experimental study of both radiation reaction and quantum effects within the limits of the BELLA PW facility. For BELLA PW (and hereon in this manuscript), we therefore choose an experimental layout in which a highly-relativistic electron beam (multi-GeV) collides with a high intensity laser pulse.

\section{Production of Monoenergetic Electron Beams in an LWFA Driven by the First Beamline}
\label{sec:ebeamsim}

\begin{table}[t]
\begin{center}
\begin{tabular}{|c||c|c|c|c|c|c|c|}
\hline
Laser energy, $U_1$ [J]& 10 & 15 & 20 & 20 & 25 & 30 & 35 \\
\hline
Laser duration, $\tau$ [fs] & 40 & 60 & 80 & 80 & 100 & 120 & 140 \\
\hline\hline
Target type &  CDW & CDW & CDW & CDW+LH & OFI & OFI & OFI \\
\hline
Stage length, $L_{plasma}$ [cm] & 7.6 & 16.6 & 24.0 & 28.0 & 36.6 & 44.8 & 80.8 \\
\hline
Plasma density, $n_0$ [$\times 10^{17}$ cm$^{-3}$] & 6.0 & 4.0 & 3.0 & 2.5 & 2.0 & 1.5 & 1.0 \\ 
\hline
Matched radius, $R_m$ [$\mu$m] & 83 & 92 & 99 & 63 & 65 & 55 & 55 \\
\hline
Ramp length, $L_{ramp}$ [cm] & 0.3 & 0.3 & 0.3 & 0.4 & 0.4 & 0.4 & 0.4\\
\hline
Dopant fraction, $f_{N/H}$ [\%] & 4 & 1 & 1 & 3 & 2 & 2 & 5\\ 
\hline
\hline
Beam charge, $Q_b$ [pC] & 9.2 & 9.1 & 10.7 & 22.6 & 30.0 & 34.0 & 10.4 \\ 
\hline
Beam energy, $\mathcal{E}_e$ [GeV] & 2.1 & 2.8 & 3.3 & 5.9 & 7.0 & 9.4 & 12.4 \\ 
\hline
Beam energy spread, $\delta \mathcal{E}_e/ \mathcal{E}_e$ [\%] & 2.0 & 2.8 & 4.5 & 3.3 & 2.0 & 3.2 & 3.7\\ 
\hline
Beam divergence, $(\theta_x\theta_y)^{1/2}$ [mrad] & 0.59 & 0.63 & 1.54 & 0.52 & 0.28 & 0.55 & 0.25 \\
\hline
\end{tabular}
\end{center}
\caption{\label{tab:LPA1bl}Summary of all the laser and plasma parameters and the corresponding final properties of the electron beams produced in an LWFA driven by the BELLA PW 1BL for different values of the laser energy and different types of plasma target (CDW, CDW+LH, and OFI). In all cases the laser spot size is $w_{\mathrm{0}}=53$ $\mu$m, the central laser wavelength is \unit[815]{nm} (\unit[$\sim800$]{nm}), and the focus position is \unit[3]{mm} downstream from the plasma entrance.}
\end{table}

Section~\ref{sec:interactiongeom} clarified that SF-QED experiments using the BELLA PW laser will require multi-GeV electron beams. As will be illustrated in Sec.~\ref{sec:expgeom}, such beams will be produced and accelerated in plasma wakefields that are driven by the pulses of BELLA PW first beamline (1BL).

Previous simulation results show that, when the full \unit[$\sim$40]{J} energy is used together with optimal plasma parameters, (quasi-)monoenergetic electron beams with energies $\gtrsim 10$ GeV and $\sim 100$ pC of charge can be produced \cite{benedetti.ieee.2018}. However, that would leave no pulse energy for 2BL, required to provide the EM field for the SF-QED interaction. Therefore, to characterize the properties of LWFA electron beams in the context of the proposed SF-QED experiments, we investigate the use of \unit[10-35]{J} laser energy in the first beamline ($U_1$), as indicated in Tab.~\ref{tab:LPA1bl}, leaving \unit[30-5]{J} of pulse energy for 2BL (the total available energy is \unit[$\sim$40]{J}). 

Simulation studies are performed using the Particle-In-Cell (PIC) code {\sc INF\&RNO} \cite{benedetti.AAC.2010, benedetti.ppcf.2017}. The longitudinal laser pulse profile is modeled as a Gaussian with a FWHM duration of \unit[$\tau=$40-140]{fs} (see Tab.~\ref{tab:LPA1bl}), depending on what is optimal for acceleration at each $U_1$. The transverse pulse profile at focus is the one corresponding to a near-field flat-top (i.e., a jinc profile) with a spot size $w_{\mathrm{0}}=53$ $\mu$m (for details on the definitions, and assumed pulse profile, see Ref.~\cite{benedetti.ieee.2018}). The central laser wavelength in the simulations is \unit[815]{nm} (\unit[$\sim800$]{nm}).

Laser and plasma parameters are chosen such that the LWFA operates in a dark current-free, mildly nonlinear regime \cite{esarey.RMP.2009}. All simulations have been performed with the quasi-static modality of INF\&RNO \cite{benedetti.aac16.2017}. Absence of high-energy particles from self-injection was verified by running fully self-consistent simulations for some of the cases. Guiding of the laser pulse, required for multi-GeV energy gains, is provided by a plasma channel with an on-axis density $n_0$ and a transverse parabolic density profile with matched radius $R_m$. The values of $n_0$ and $R_m$ considered in this study are the ones experimentally obtainable with a capillary discharge waveguide (CDW) \cite{spence.pre.2000} with a radius of $R_{cap}=300$ $\mu$m (using capillaries with a smaller radius might result in damage of the structure by the wings of the pulse) or, in the cases corresponding to lower densities ($n_0\lesssim 2.5\times 10^{17}$ cm$^{-3}$) and smaller values of the matched radius ($R_m\lesssim 65$ $\mu$m), by enhancing the performance of the CDW with the laser-heater technique (CDW+LH) \cite{bobrova.pop.2013, gonsalves.prl.2019, pieronek.pop.2020}, or, finally, by employing an optical-field-ionized (OFI) channel \cite{shalloo.prab.2019, miao.prl.2020}. Note that in a CDW the values of $n_0$ and $R_m$ are related (e.g., $R_m[\mu \hbox{m}]\simeq 7.5 R_{cap}[\mu \hbox{m}])^{1/2}/(n_0[10^{17}\hbox{cm}^{-3}])^{1/4}$). Using the LH technique or an OFI channel allows, in principle, for independent control of these parameters, but will require additional laser pulses.
The plasma profile is longitudinally uniform with entrance and exit ramps of length $L_{ramp}$ (with a square root-like profile), and the total plasma length is $L_{plasma}$.

The electron beam is produced by ionization-induced injection when the laser pulse enters the LWFA stage. This is achieved by concentrating a small amount of a high-Z dopant gas (Nitrogen in this case) within the plasma entrance ramp \cite{gonsalves.pop.2020}. The fraction of Nitrogen atoms with respect to the background Hydrogen atoms is denoted as $f_{N/H}$ in Tab.~\ref{tab:LPA1bl}.       

For every value of laser pulse energy $U_1$, the laser pulse duration $\tau$, the on-axis plasma density $n_0$, and the matched radius $R_m$ were chosen to guarantee, compatibility with the chosen plasma target type, sufficient laser guiding and, hence, reasonably stable wake properties over the desired acceleration length. For each case the length of the up-ramp (plasma entrance) and concentration of the dopant $f_{N/H}$ were adjusted to control beam charge and energy spread. For all the cases, the laser focus position is \unit[3]{mm} downstream from the plasma entrance. The final properties of the electron beams, together with all relevant laser and plasma parameters, are summarized in Tab.~\ref{tab:LPA1bl}. Further optimizations, e.g., in beam charge and/or energy, are possible. 

Previous experiments on BELLA PW produced up to \unit[7.8]{GeV}, \unit[5]{pC} electron beams using \unit[30]{J} of laser energy in a \unit[20]{cm}-long plasma \cite{gonsalves.prl.2019}. The results of Tab.~\ref{tab:LPA1bl} show that multi-GeV (\unit[$\mathcal{E}_e=2.1-12.4$]{GeV}), quasi-monochromatic (\unit[$\delta \mathcal{E}_e/\mathcal{E}_e=2-4.5$]{\%}), low-divergence (\unit[$(\theta_x\theta_y)^{1/2}=0.25-1.54$]{mrad}) electron beams with \unit[$Q_b=9-34$]{pC} of charge are within reach given the current experimental capabilities of BELLA PW, or can become available in the near future. Critical for the production of these high-quality, high-energy beams is the implementation of a guiding technology that is able to produce strong guiding plasma structures in low-density plasmas, which is already part of current the experimental program of the BELLA PW facility (see Sec.~\ref{sec:plannedexp}).  

\section{SF-QED Simulation Results}
\label{sec:SFQEDsimres}

In this section, we present {\sc ptarmigan} \cite{blackburn.njp.2021,Blackburn:2021cuq,parmigansourcecode} simulation results of an electron beam interacting with a counter-propagating laser pulse (geometry chosen in Sec.~\ref{sec:interactiongeom}). {\sc ptarmigan} is a single particle code that models interactions using classical dynamics of charged particles and SF-QED processes, taking into account the angular distributions of secondary particles produced in either Compton or Breit-Wheeler processes. The SF-QED processes are modeled in the framework of the local constant field approximation (LCFA), which is valid for the laser intensities and particle energies considered below. 

The laser pulse is specified using the paraxial solution for the fields given in Ref. \cite{salamin.apb.2007} with terms up to the fourth order in the diffraction angle in the Gaussian beam, which is then multiplied by a temporal envelope function. The laser pulse is linearly polarized and defined by the peak value of the field strength ($a_0$), wavelength ($\lambda$), waist ($w_0$, defined as the radius where the intensity falls to $1/e^2$ of its maximum value) and pulse duration ($\tau$, which is full width at half maximum). The electron beam is defined by the mean Lorentz factor ($\gamma$), energy spread ($\delta \mathcal{E}_e/ \mathcal{E}_e$), Gaussian transverse ($w_e$) and longitudinal ($l_e$) beam charge distribution  and divergence ($(\theta_x\theta_y)^{1/2}$, normally distributed). Experimentally achievable electron energies (see Sec.~\ref{sec:ebeamsim}, Tab.~\ref{tab:LPA1bl}) and laser intensities (see Sec.~\ref{sec:expgeom}) were used as input. 
 
 %parameters simulated in  and the laser pulse intensities ($a_0$, where $a_0\simeq0.855\times 10^9(I$[W/cm$^2$])$^{1/2}\lambda$[um])) discussed in Sec.~\ref{fig:LayoutChamber},

\begin{figure}[htb!]
    \centering
    \includegraphics[width=0.8\columnwidth]{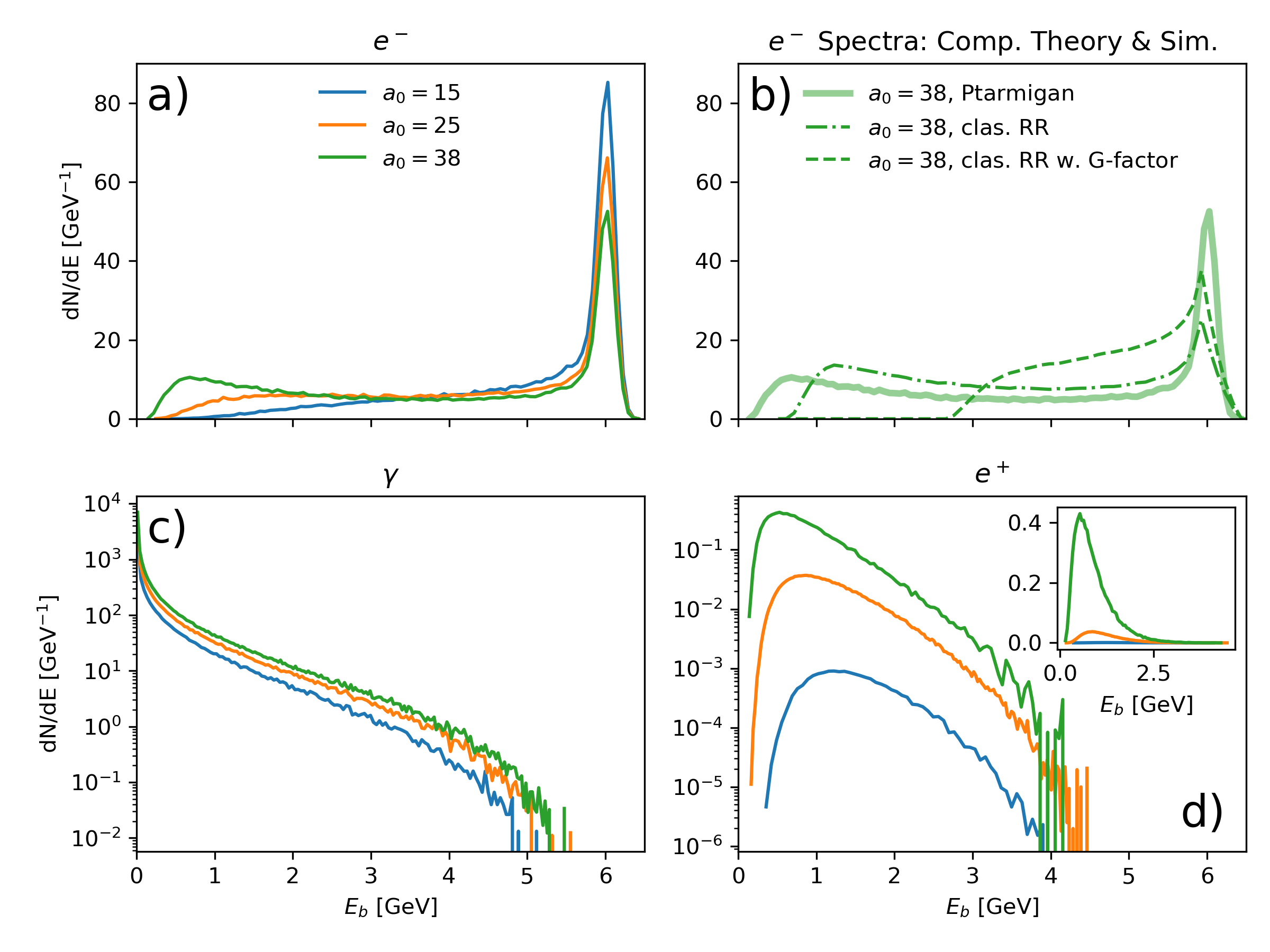}
    \caption{Energy spectra of (a) electrons $e^{-}$, (c) photons $\gamma$ and (d) positrons $e^{+}$ after the interaction of a \unit[$\mathcal{E}_e=5.9$]{GeV}, \unit[$Q_b=22.6$]{pC} electron beam with a counter-propagating laser pulse with $a_0=15$ (blue), $a_0=25$ (orange) and $a_0=38$ (green) as simulated using {\sc ptarmigan}. Note that panels (c) and (d) are on a logarithmic vertical scale and that the inset in panel (d) shows the same curves as on panel (d), but on a linear scale.  The input electron beam has relative energy spread of \unit[$\delta \mathcal{E}_e/ \mathcal{E}_e=3.3$]{\%}. The legend in panel (a) applies also to panels (c) and (d). Panel (b) compares the predicted electron energy spectra for $a_0=38$ from {\sc ptarmigan} simulation results (solid line), classical theory including radiation reaction (dash-dotted line) and classical theory including radiation reaction using the G-factor (dashed line).}
    \label{fig:Spectra}
\end{figure}

For example, Fig.~\ref{fig:Spectra} compares the energy spectra of electrons (a), photons (c) and positrons (d) after the interaction of a \unit[$\mathcal{E}_e=$5.9]{GeV}, \unit[$Q_b=22.6$]{pC} electron beam with a laser pulse of $a_0=15$ (blue), $a_0=25$ (orange), and $a_0=38$ (green). The laser pulse is assumed to have a wavelength of \unit[$\lambda=800$]{nm}, \unit[$\tau=30$]{fs} duration, and a focal spot size of \unit[$w_{\mathrm{0}}=2$]{um}. The initial electron energy spectrum is quasi-monoenergetic with parameters according to Tab.~\ref{tab:LPA1bl} and \unit[$w_e=l_e=2$]{um} transverse and longitudinal beam sizes at the interaction point.

In the examples of Fig.~\ref{fig:Spectra} (a), electrons loose a significant amount of energy due to radiation (e.g., 37\% of initial beam energy for $a_0=38$) and develop a broad energy distribution. For $a_0=38$, a distinct second maximum appears around \unit[0.5]{GeV} and each incoming electron emits on average approximately eight photons, with a broad distribution peaked towards zero (see Fig.~\ref{fig:Spectra} (c)). Less photons are being emitted for lower values of $a_0$, which can be observed from the photon spectra (see Fig.~\ref{fig:Spectra}(c)) and the disappearance of the second maximum in the electron spectra for $a_0=15$ and 25 (see Fig.~\ref{fig:Spectra}(a)).

A small fraction of these photons decayed into electron-positron pairs (around one pair per 130 initial electrons for $a_0=38$) via the multi-photon Breit-Wheeler process, where "multi-photon" refers to the interaction with the fixed classical background field \cite{piazza.rmp.2012,gonoskov.rmp.2022}. The positron energy spectra (see Fig.~\ref{fig:Spectra}(d)) have a maximum at \unit[$\sim$0.5]{GeV} ($\gamma\sim 1000$), which is mainly formed by photons, whose radiation length is about the length of the laser pulse \cite{bulanov.pra.2013}. Simulation results predict a strong increase in electron-positron pair production for higher values of $a_0$. For example, for $a_0=15$ the curve is indistinguishable from zero on a linear scale also showing the other curves (see inset of Fig.~\ref{fig:Spectra} (d)). Increasing $a_0$ from 25 to 38 increases the number of electron-positron pairs by one order of magnitude.

Figure~\ref{fig:Spectra}(b) demonstrates that experiments on BELLA PW will allow to compare results to different interaction descriptions, e.g., obtained from the classical or the SF-QED framework. To illustrate that, we show three electron spectra after the interaction of a \unit[$\mathcal{E}_e=5.9$]{GeV}, \unit[$Q_b=22.6$]{pC} electron beam with a counter-propagating laser pulse with $a_0=38$ obtained using different frameworks, calculated using the same input parameters. The green solid line is the same as in Fig.~\ref{fig:Spectra}(a), and represents the simulation result obtained with {\sc ptarmigan}  predicting \unit[37]{\%} energy loss. The dashed-dotted line was obtained from the solution of the Landau-Lifshitz equation  and predicts \unit[64]{\%} energy loss. The dashed line was obtained from the solution of the "modified" Landau-Lifshitz equation \cite{gonoskov.rmp.2022,Blackburn:2021cuq}  and predicts \unit[51]{\%} energy loss. In the "modified" Landau-Lifshitz equation an additional factor, $G(\chi)$, was introduced before the radiation reaction force to account for the classical overestimation of the amount of radiation emitted by an electron in strong fields. This factor $G(\chi)$ gives the ratio between the instantaneous radiation powers predicted by QED and by the classical theory. Both classical calculations predict higher levels of energy loss than the SF-QED result. Moreover, the shape of the spectra is different, which will help to identify the limits of the applicability of each theory when compared to experimental results. 

\begin{figure}[htb!]
    \centering
    \includegraphics[width=\columnwidth]{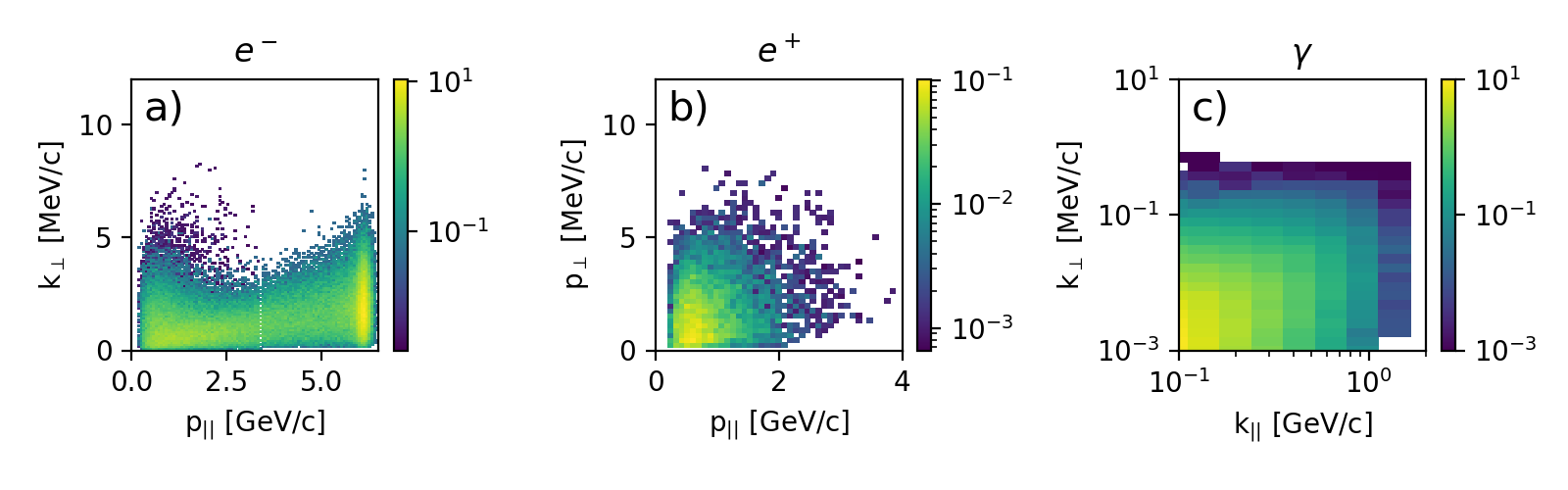}
    \caption{Phase space of electrons $e^{-}$ (a), positrons $e^{+}$ (b) and photons $\gamma$ (c) after the head-on collision of a \unit[$\mathcal{E}_e=5.9$]{GeV} electron beam with a counter-propagating laser pulse pulse with $a_0=38$. Note the logarithmic color scales, which are in arbitrary units and normalised to the pixel area.}
    \label{fig:FinalDistr}
\end{figure}

Figure~\ref{fig:FinalDistr} shows the phase space for the electrons (a), positrons (b), and photons (c) for the case with $a_0=38$, after the interaction. % The electrons develop broad distribution spanning from the initial energy all the way to low energies (Fig. \ref{fig:FinalDistr} (a)), which is the indication of a significant energy transfer to the photons (Fig. \ref{fig:FinalDistr} (c)) mentioned above. The distribution of positrons is more confined near the maximum at \unit[$\sim$0.5]{GeV} (Fig. \ref{fig:FinalDistr} (b)). 
All three distributions demonstrate that the particles are moving in the forward direction inside a narrow cone, mainly determined by $1/\gamma$. All beams overlap with each other in space. These distributions informed the design of the experimental diagnostics.

Figures~\ref{fig:Spectra} and \ref{fig:FinalDistr} show a typical predicted experimental outcome of an electron beam interacting with a counter-propagating laser pulse. For $a_0=38$, the interaction is well into the quantum regime ($\chi_{e,max}=2.1$). However, $\chi_{e,max}$ could be lowered by, e.g., adjusting $a_0$ (e.g. by lowering the laser pulse energy) in order to study the transition of the radiation reaction from the quantum to the classical regime.

\begin{figure}[htb!]
    \centering
    \includegraphics[width=0.9\columnwidth]{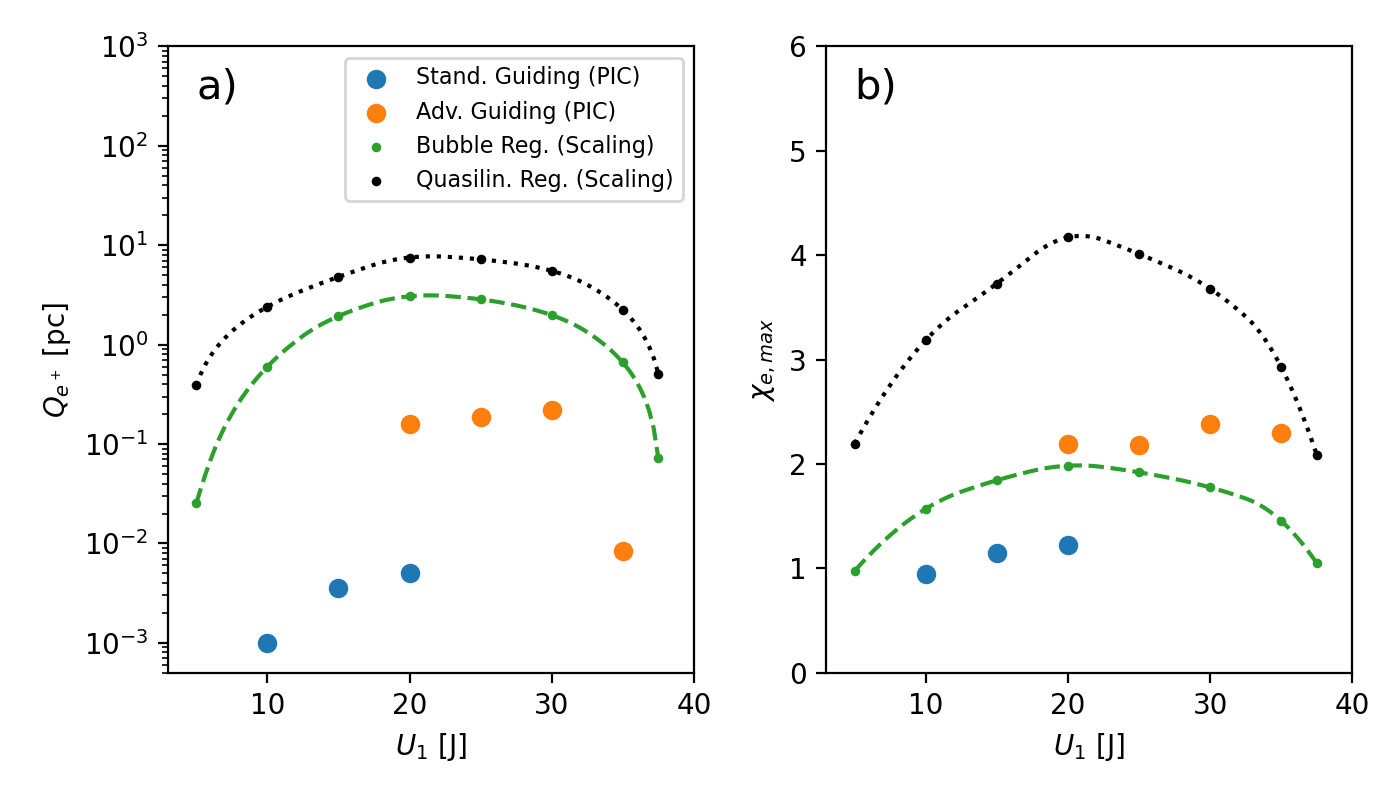}
    \caption{{\sc ptarmigan} simulation results on the produced positron charge $Q_{e^+}$ (a) and maximum $\chi_{e,max}$ (b) reached during the head-on collision of a laser pulse with energy $U_1$ and an electron beam of the corresponding energy as listed in Tab.~\ref{tab:LPA1bl} (PIC) or using the LWFA electron beam scalings described in the appendix (Scaling) as input. Blue points show the standard experimental configuration in which the LWFA uses capillary discharge waveguide (CDW) guiding structure (see simulated electron beam parameters in Tab.~\ref{tab:LPA1bl}). Orange points show what could be achievable when the LWFA uses advanced guiding structures (LH and OFI, see Tab.~\ref{tab:LPA1bl}). For $U_1=10, 15, 20, 25, 30, 35$ and $37.5$ the $a_\mathrm{0}$ values are $50, 46, 42, 38, 32, 26, 19$ and $13$, respectively. Black and green dots show the theoretically achievable value when estimating the LWFA parameters in the bubble and quasi-linear regime from the scalings described in the appendix, respectively. Black dotted and green dashed lines show quadratic interpolations of the correspondingly colored data points. Note the logarithmic vertical scale for the left plot. Panel b) shares the legend shown in a).}
    \label{fig:charge}
\end{figure}

Figure \ref{fig:charge} summarizes the produced integrated positron beam charge ($Q_{e^+}$, see blue and orange dots in Fig.~\ref{fig:charge}(a)) and maximum $\chi_{e,max}$ (see blue and orange dots in Fig.~\ref{fig:charge}(b)) as a function of the LWFA laser energy ($U_1$) for the PIC simulations presented in Tab.~\ref{tab:LPA1bl}. The remaining laser pulse energy (\unit[40]{J}$-U_1$ and up to \unit[$\lesssim 20$]{J}), is used in the second pulse for scattering. 

However, to illustrate the potential of the BELLA PW facility when electron beams are further optimized, Fig.~\ref{fig:charge} also shows two sets of curves (black dotted and dashed green) based on idealized LWFA stages operating in the quasi-linear and bubble regimes, respectively. Details for these idealized LWFA stages are discussed in the Appendix. 

Figure \ref{fig:charge} (a) shows that the positron charge ($Q_{e^+}$) obtainable in the idealized quasi-linear case is around \unit[7]{pC}, whereas PIC simulation results predict \unit[$\sim$200]{fC}, almost a two order of magnitude difference. That increase is a result of the three times higher electron beam charge and \unit[30]{\%} higher beam energy in the quasi-linear case compared to the PIC simulations. Positron production may therefore be increased by optimization of the LWFA, e.g. by optimizing, injection efficiency, beam loading, dephasing, and depletion. We note that the parameter $\chi_{e,max}$ is larger than unity for all cases with an optimized plasma channel.

As emphasized in previous experimental and theoretical studies (see, e.g., Ref.~\cite{gonoskov.rmp.2022} and Fig.~\ref{fig:Spectra}(b)), the amount of energy loss and the form of the final electron spectrum play a crucial role in determining the interaction regime, i.e., whether the interaction requires a quantum description or can be described in the framework of classical electrodynamics. Experimental results will therefore be compared to theoretical and simulation results to obtain further insight.

The simulations described in this Section were performed using LCFA, which is a standard approach when considering SF-QED effects. However, it was shown recently that the LCFA fails at moderate values of $a_0$ ($\sim 1-10$) by overestimating the number of low-energy photons generated \cite{harvey.pra.2015,dinu.prl.2016,dipiazza.pra.2018,blackburn.pop.2018}. Several solutions were proposed \cite{dipiazza.pra.2018,ilderton.pra.2019,blackburn.pra.2020}, including using the locally monochromatic approximation (LMA) instead of the LCFA \cite{blackburn.njp.2021,heinzl.pra.2020}. The main difference between the two is that LMA includes interference effects at the scale of the laser wavelength. For the values of electron beam energy which were used in the simulations presented above, the results indicate that at $a_0=1-10$, the LCFA overestimates the number of photons by approximately \unit[10]{\%}, when compared to the LMA result. Thus, the experiments at the BELLA PW facility can provide invaluable input into determining the applicability range of different approximations. 

Another challenge of SF-QED theory is the description of multi-staged processes with identification of possible interference effects. Usually these processes are treated independently for each stage. The possible observation of electron-positron pair production in laser electron beam collision opens the possibility to study the ``trident'' ($e^-\rightarrow e^-e^+e^-$) process and to identify whether treating each stage as independent is justified (see, e.g., Refs.~\cite{mackenroth.prd.2018,dinu.prd.2018,king.prd.2018}).

\section{Experimental Layout for SF-QED Experiments}
\label{sec:expgeom}

\begin{figure}[htb!]
    \centering
    \includegraphics[width=\columnwidth]{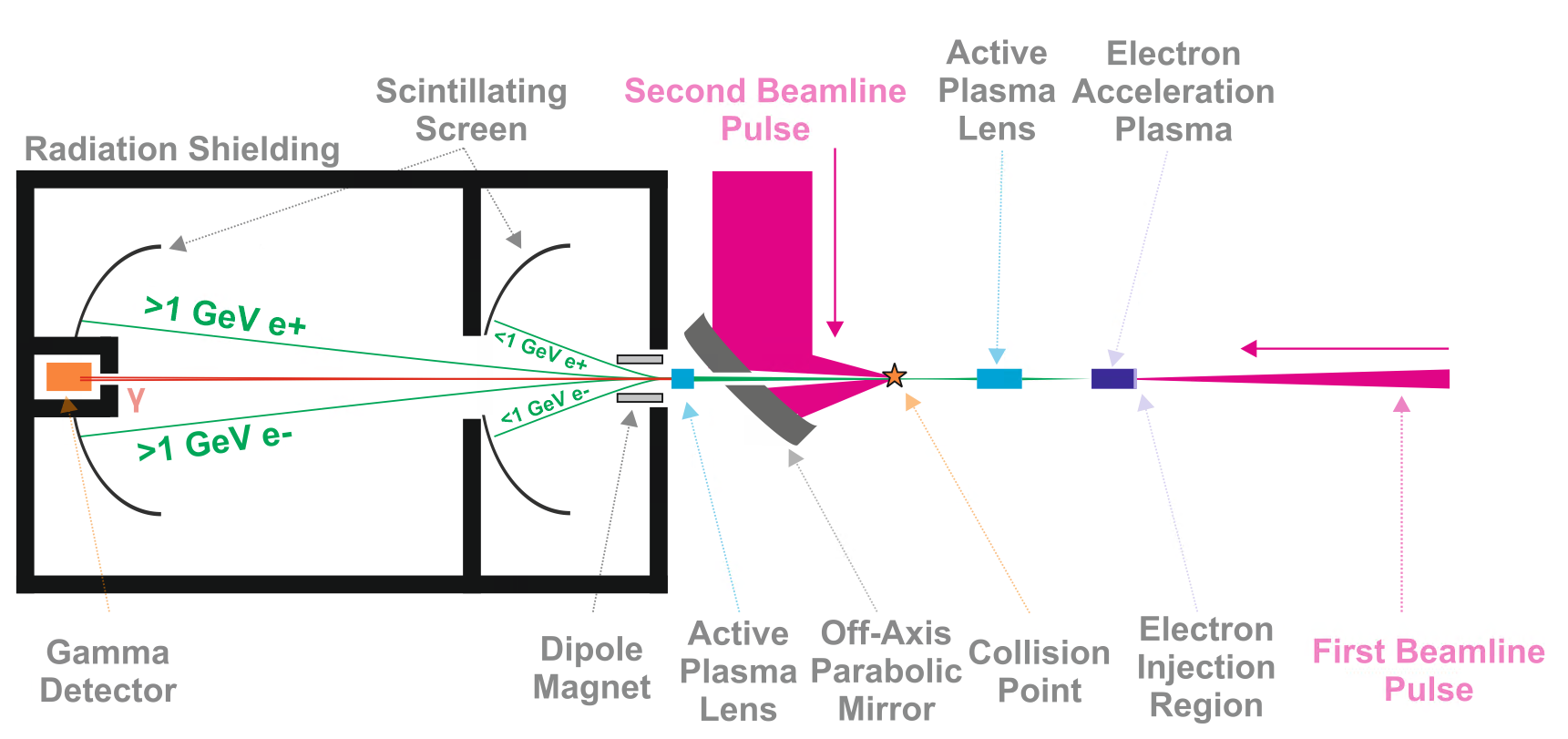}
    \caption{Schematic top-view layout of a head-on SF-QED collision experiment in the BELLA Petawatt interaction chamber. }
    \label{fig:LayoutChamber}
\end{figure}

Figure~\ref{fig:LayoutChamber} shows a top-view, schematic experimental setup for SF-QED experiments at the BELLA PW facility, is based on geometry 1) discussed in Sec.~\ref{sec:interactiongeom}, and is similar to the layouts used in previous and planned experiments from other facilities (see e.g., Refs.~\cite{bula.prl.1996,cole.prx.2018,poder.prx.2018,abramowicz.arxiv.2021}). The first beamline (1BL) pulse arrives from the right and is focused around the entrance of the plasma used for electron acceleration. As discussed in Sec.~\ref{sec:ebeamsim}, this pulse is used to both produce electron beams via ionisation injection and to drive plasma wakefields with $\sim$GV/m field amplitudes. 

%Previous experiments using the BELLA PW laser facility with similar experimental setups showed that \unit[16]{J} of laser energy can accelerate \unit[6]{pC} of electrons up to \unit[4.2]{GeV} energies in a \unit[9]{cm}-long capillary plasma waveguide or \unit[$\sim$30]{J} laser pulse energy to \unit[7.8]{GeV} energies in a \unit[20]{cm}-long capillary plasma waveguide. Achievable electron beam parameters as a function of laser pulse energy in the context of the proposed SF-QED experiments were discussed in Sec.~\ref{sec:ebeamsim}.

After acceleration, the multi-GeV electron beam is refocused using an active plasma lens (APL) (as shown in the schematic of Fig.~\ref{fig:LayoutChamber}), which is a compact alternative ($\sim$cm-length at GeV energies) to a lens based on a magnetic quadrupole triplet. Using an APL has at least two advantages: 1) it provides control of the transverse electron beam spot size at the collision point. This is desirable as an approximately equal electron beam and laser transverse pulse spot size maximises the interaction cross-section; 2) APL focusing is chromatic, which means that the focal position depends on the electron energy. This allows the selection of electron energy by maximizing the interaction probability at the plane of the laser focus (which is also the collision point). 

Figure~\ref{fig:LayoutChamber} also illustrates that the collimated second beamline pulse arrives at an angle of \unit[90]{degrees} with respect to the first beamline. After entering the chamber, it is focused with a short focal length (f/1-f/2.5) \unit[45]{degree}, OAP mirror with a hole. The hole is required to allow the counter-propagating beams to reach the shielded diagnostics area. The size of the hole will be negligible compared to the total pulse size and will therefore not decrease 2BL pulse energy significantly. The near field profile of the pulse is top-hat shaped and a jinc profile provides an approximate description of the transverse pulse shape at focus.

We first discuss the ideal scenario, in which the facility is upgraded, such that the full aperture pulse can be transported and focused inside the target chamber. Using, e.g., a f/1-f/2.5 off-axis parabolic mirror would provide an intensity of $a_{\mathrm{0}}= 201.6 (\lambda / w_0)  ( U_1[\hbox{J}] / \tau[\hbox{fs}] )^{1/2}$ up to $\sim80$, where $w_{\mathrm{0}}$ is the pulse spot size at focus (which is also the collision point, \unit[$w_{\mathrm{0}}=$1-3]{um}), $U_1$ is the laser pulse energy (e.g., \unit[20]{J}), and $\tau$ is the pulse length (\unit[$\tau$=30]{fs}). An achievable Strehl ratio of 0.8 was assumed. Without the installation of a new target chamber, the maximum beam size that can be transported into the chamber is of a diameter of \unit[15]{cm}. Reducing the beam diameter from 20 to \unit[15]{cm} reduces pulse energy %by \unit[$\sim 44$]{\%} 
and increases the focal spot size ($w_{\mathrm{0}}$) allowing for maximum laser strength of $a_{\mathrm{0}} \approx 40$. After the interaction, the laser pulse will be dumped and its energy dispersed inside the target chamber. For experiments that aim at measuring the transition between regimes, the pulse energy in the 2BL (and therefore $a_0$) can be lowered by, e.g., using a waveplate-polarizer setup before pulse compression. Since it is expected to be very challenging to produce $a_\mathrm{0}=80$ experimentally, a maximum of $a_\mathrm{0}=50$ was used in Sec.~\ref{sec:SFQEDsimres}.

Electron, positron, and photon beam diagnostics will be installed downstream the collision point and downstream the OAP in the propagation direction of the 1BL pulse. A \unit[$\sim1$]{T}, tens-of-centimeter long dipole magnetic field disperses the charged particles depending on their energy and charge. Charged particles below \unit[1]{GeV} energy are measured on a scintillating screen inside the target chamber, the ones above \unit[1]{GeV} are measured on a separate scintillating screen several meters further downstream. At this longitudinal position, $>$GeV charged particles are deflected far-enough off-axis (given their $\sim$mrad divergence) for high-energy photons to be measured on-axis using a shielded radiation detector. When desired, a second APL may be used to image a selected particle energy onto the plane of the spectrometer screen. 

It is clear that the experimental diagnostics design and implementation for strong-field QED experiments is challenging and will require stable electron beams, advanced beam and pulse diagnostics to monitor shot-to-shot alignment, calibrated charged particle and photon measurements together with sophisticated radiation shielding, due to the large discrepancies between the amount of particles, their wide energy range, the required measurement accuracy, and the large expected background.

\section{Summary \& Conclusions}
\label{sec:Conclusions}

In this paper, we discussed the implementation and scientific reach of future SF-QED experiments at the BELLA PW laser facility. The experimental capabilities of the facility are unique because of the \unit[1]{Hz} repetition rate of the laser system, enabling parameter optimization and the possibility to obtain good statistics for a wide set of parameters.

Experiments will use the \unit[40]{J} BELLA PW laser energy in two high-intensity pulses. The first pulse will produce and accelerate electrons to multi-GeV energies in a laser-driven plasma accelerator (LWFA). The remaining pulse energy is in the second pulse and provides the EM field for the scattering with the electron beam. We presented simulation results showing that experiments using the BELLA PW laser can 1) reach a maximum nonlinear quantum parameter $\chi_{e,max}$ of up to 4; 2) provide access to the SF-QED radiation reaction and quantum interaction regimes, and 3) produce positron beams with $\sim$ mrad divergence and a charge of fC to pC.

These experiments may allow testing of the theoretical SF-QED models and validations of the approximations used in theory and simulations. Experimentally obtained spectra should also allow to determine in which regime the interaction occurred, i.e., whether it entered the quantum regime, or can still be described in the framework of classical electrodynamics. Additionally experiments will allow to study and evaluate whether SF-QED interactions could be used as a source of positrons for future applications, such as, e.g., a next generation electron-positron collider.

%Based on our simulations of the high-intensity laser pulse collision with a multi GeV-class electron beam, one can expect the experiments at BELLA PW facility to: 1) demonstrate electron beam depletion with around 30\% energy loss; 2) produce a copious amount high energy photons (with almost all energy lost by electrons to be carried away by photons); 3) produce electron-positron pairs by the multi-photon Breit-Wheeler effect as well as 4) provide an opportunity to study multi-staged processes, including `trident' and double Compton effects. The analysis of the energy partition and the distributions of particles in the final state should allow to determine in which regime the interaction occurred, i.e., whether it entered the quantum regime, or can still be described in the framework of classical electrodynamics. 
\section{Data Management Plan}
The datasets generated during and/or analysed during the current study are available from the corresponding author on reasonable request. Simulation input is available from Stepan Bulanov (sbulanov@lbl.gov) on reasonable request.

\section{Acknowledgements}
This work was supported by the Director, Office of Science, Office of High Energy Physics, of the U.S. Department of Energy, under Contract No. DE-AC02-05CH11231, and used the computational facilities at the National Energy Research Scientific Computing Center (NERSC). We acknowledge helpful discussions with T. Blackburn regarding the {\sc ptarmigan} code. The contributions from W.~P. Leemans were made prior to his 2019 departure from LBNL to DESY.

%\appendix

\section*{Appendix -- Idealized LWFA Stages in the Quasi-Linear and Bubble Regime}
\label{sec:Appendix}
In the following, we describe the details of the idealized LWFA stages discussed in Sec.~\ref{sec:SFQEDsimres}. 

For the idealized stage in the quasi-linear regime (see black dotted lines in  Fig.~\ref{fig:charge}) we considered an LWFA driven by a super-matched (see Ref.~\cite{benedetti.pre.2015} for details on the definition) laser pulse with $a_0=1.6$, $k_p w_0=4$, and $k_p c T_{fwhm}=2.12$ (Gaussian longitudinal profile). Here $k_p=(4\pi n_0 e^2/mc^2)^{1/2}$ is the plasma wavenumber. The central laser wavelength is \unit[800]{nm}. The operational density is specified once the laser energy is specified and is given by $n_0[\hbox{cm}^{-3}]\simeq 7.14\times 10^{17} (U_1 [J])^{-2/3}$. To guide the laser a plasma with a parabolic transverse density profile, $R_m=w_0$ is used. 

For the stage operating in the bubble regime (see green dashed lines in Fig.~\ref{fig:charge}), laser driver is bi-Gaussian and its intensity is such that $a_0=4.5$, furthermore laser focal spot size $w_{\mathrm{0}}$ and pulse length $\tau$ are chosen according to the theory in Ref.~\cite{lu.prstab.2007} (i.e., $k_p w_0=2\sqrt{a_0}$, and $cT_{fwhm}=(2/3)w_0$), and the central laser wavelength is \unit[800]{nm}. As before, the operational density of the stage is specified once the laser energy is specified and is given by $n_0[\hbox{cm}^{-3}]\simeq 7.02\times 10^{18} (U_1 [J])^{-2/3}$. Note that, for a given laser energy, and for the parameters considered here, the density of a stage operating in the quasi-linear regime is about an order of magnitude lower compared to the one of a stage operating in the bubble regime. Due to the longer dephasing and depletion lengths at lower densities, the energy gain provided by a quasi-linear stage is generally larger than that provided by a stage operating in the bubble regime.

In both, the quasi-linear and bubble stages, the initial electron beam is chosen to experience $\sim$75\% of the maximum accelerating field (for the stage in the bubble regime the maximum field is obtained with a linear extrapolation of the longitudinal wake to the back of the bubble), and the current profile is such that the longitudinal wakefield in the beam region is initially flat (i.e., strongly beamloaded stages). The charge of the electron beam is $Q_b[\hbox{pC}]\simeq 37 (U_1 [J])^{1/3}$ in the quasi-linear stage, and $Q_b[\hbox{pC}]\simeq 139 (U_1 [J])^{1/3}$ for the bubble case.

\section{Author Contribution Statement}

M. Turner lead the experimental design effort and described worked on the installation and commissioning of BELLA PW second beamline together with A.~J. Gonsalves and K. Nakamura . S.~S. Bulanov performed the electron beam-laser pulse interaction simulations for this manuscript and provided the theoretical SF-QED paper discussion. C. Benedetti simulated and optimized the laser wakefield accelerated electron beams. W.~P. Leemans was involved in the initial design and realisation of the BELLA PW second beamline project. J. van Tilborg,  C.~B. Schroeder, C.~G.~R. Geddes and E. Esarey provided input at all stages of the manuscript preparation and coordinated and supervised the efforts.

\bibliographystyle{ieeetr}
\bibliography{main}

\end{document}